\documentstyle[aps,twocolumn]{revtex}

\begin{document}
\draft
\title{Operational Theory of Homodyne Detection}
\author{Konrad Banaszek\cite{uw}}
\address{Optics Section, Blackett Laboratory,
Imperial College, Prince Consort Road,\\
London SW7 2BZ, United Kingdom}
\author{ Krzysztof W\'odkiewicz\cite{uw}}
\address{Center for Advanced Studies and 
Department of Physics and Astronomy,\\
University of New Mexico, Albuquerque NM 87131, USA}

\date{September 27, 1996}
\maketitle

\begin{abstract}
We discuss a balanced homodyne detection scheme with imperfect
detectors in the framework of the operational approach to quantum
measurement. We show that a realistic homodyne measurement is
described by a family of operational observables that depends on
the experimental setup, 
rather than a single field quadrature operator. 
We find an explicit form of this family, which fully
characterizes the experimental device and is independent of a specific
state of the measured system.   
We also derive  operational homodyne observables for the setup with
a random phase, which has been recently applied in an ultrafast
measurement of the photon statistics of a pulsed diode laser. 
The operational formulation directly gives the relation
between the detected noise and the intrinsic quantum fluctuations of
the measured field. We demonstrate this on two examples: the
operational uncertainty relation for the field quadratures, and the
homodyne detection of suppressed fluctuations in photon statistics. 
\end{abstract}

\section{Introduction}
Homodyne detection is a well known 
technique in detecting phase--dependent
properties of optical radiation. 
In quantum optics it has been
widely  used in studies and applications
of squeezed light \cite{squeezing}.
A statistical distribution of the outcomes of a homodyne detector
has recently found  novel
applications in the measurement of the quantum state of light 
via optical homodyne tomography \cite{tomography} and the direct probing of
 quantum phase space by photon counting \cite{banaszek}.
The phase--sensitivity of  homodyne detection is achieved by 
performing a
superposition of the signal field with a coherent
local oscillator by means of a beamsplitter \cite{YuenShapCQOIV}. 
It was an important observation \cite{TwoPapersOL} 
that in a balanced scheme, with a 50\%:50\% beamsplitter,
the local oscillator noise
can be cancelled by subtracting the photocurrents of the
detectors facing two outgoing beams. Then, in the limit 
of a classical local oscillator, the statistics
of difference photocounts is simply a signal quadrature
distribution rescaled by the amplitude of the local oscillator
\cite{BrauPRA90,VogeGrabPRA93}. 
Therefore, balanced homodyne detection is an optical realization
 of an abstract quantum mechanical measurement 
of the field quadratures described by a quantum observable
$\hat{x}_{\theta}$. The statistical outcomes of an ideal 
measurement of  $\hat{x}_{\theta}|{x}_{\theta}\rangle =
{x}_{\theta} |{x}_{\theta}\rangle $, are described by the spectral
measure 
\begin{equation}
p({x}_{\theta}) = \langle\, 
| {x}_{\theta}\rangle \langle {x}_{\theta}|
\,\rangle.
\end{equation}
Although the spectral measure contains all 
the relevant statistical information about the 
homodyne measurement, it corresponds  to a 
quantity that is measured by an ideal noise--free detector. 
Due to this property $\hat{x}_{\theta}$ 
will be called an {\it intrinsic homodyne quantum observable}.

However analysis of the homodyne setup with imperfect detectors 
\cite{VogeGrabPRA93} shows that the relation between the statistics of
the difference counts  and the quadrature spectral distribution is in
fact more 
complicated. The distribution measured in a real experiment 
is smoothed by a convolution with a Gaussian function of width
dependent on the detector efficiency. Consequently, 
realistic homodyne detection cannot be straightforwardly
interpreted as a measurement of the intrinsic field quadratures
$\hat{x}_{\theta}$. 

A recent experimental application of homodyne detection to the
reconstruction of 
the photon number distribution of a weak field from a pulsed diode
laser \cite{MunrBoggPRA95} has shown that a homodyne setup with the
fluctuating phase $\theta$ is a powerful tool in measuring
phase--insensitive properties of light. However, it is not possible 
to associate with this setup any spectral measure even in the case of 
perfect detectors. Therefore, homodyne detection with the random
phase cannot be described in terms of measuring any intrinsic quantum
observable. 
 
It is the purpose of this paper to 
show that homodyne detection provides an
interesting and nontrivial example 
of a realistic quantum measurement
leading to  operational quantum observables,
i.e., to quantum operators  that  depend
on properties of a specific experimental setup used in the homodyne
detection.  
In particular, these operational observables will depend on the
detector losses described by a quantum efficiency $\eta$ and on the
phase $\theta$ of the local oscillator used to probe the signal
field. Such operational observables provide a natural link between  
the quantum formalism and raw data recorded in a realistic homodyne
experiment. 

General features of the operational 
approach, with references to earlier
literature are given in 
\cite{EnglWodkPRA95}. The main conclusions of this
approach, if applied to the  homodyne measurement, can be summarized
as follows. 
A quantity delivered by the homodyne  
experiment is a propensity density
$\text{Pr} (a)$ of a certain classical 
variable $a$. This density 
is given by an expectation value
of an $a$-dependent positive operator valued measure (POVM), denoted
by $\hat{\cal H} (a)$:  
\begin{equation}
\text{Pr} (a) = \langle \hat{\cal H} (a) \rangle.
\end{equation}
Thus, the POVM given by $\hat{\cal H} (a)$ corresponds to 
a realistic homodyne detection and  is the mathematical
representation of the device dependent measurement. In one way of
looking at  quantum measurements, the emphasis is put 
on the construction and properties of
such POVMs. In such an approach, 
in realistic homodyne detection, 
the spectral decomposition  
 $ \text{d} {x}_{\theta} | {x}_{\theta}\rangle 
\langle {x}_{\theta}|$ of the intrinsic 
observable $\hat{x}_{\theta}$, 
is effectively replaced by the 
POVM $ \text{d} a \hat{\cal H} (a)$. 
Consequently the moments of $\text{Pr} (a)$
can be represented as
\begin{equation}
\label{OQOdef}
\overline{a^n} 
=  \int \text{d}a\, a^{n}\text{Pr} (a) = \left\langle
\mbox{$\hat{x}^{(n)}_{\theta}$}_{\cal H} \right\rangle, 
\end{equation}
defining in this way a family of {\it operational homodyne quantum
observables} 
\begin{equation}
\label{operadef}
\mbox{$\hat{x}^{(n)}_{\theta}$}_{\cal H} 
= \int \mbox{d}a\, a^n \hat{\cal H}(a),
\end{equation}
where the index ${\cal H}$ stands for 
the homodyne detection scheme associated with 
the given POVM. This family characterizes
the experimental device and is independent on a specific state of
the measured system.

In this paper we derive and discuss 
the family of operational observables 
for balanced homodyne 
detection with imperfect photodetectors. 
We show that for balanced homodyne
detection an exact reconstruction of 
the POVM $\hat{\cal H} (a)$  and of 
the corresponding operational quantum
quadratures $\mbox{$\hat{x}^{(n)}_{\theta}$}_{\cal H} $
can be performed. Thus, homodyne 
detection provides a nontrivial
measurement scheme for which an 
exact derivation of the corresponding POVM
and the  operational observables is 
possible. The interest in construction of this 
operational algebra is due to 
the fact that the number of physical
examples where the operational 
description can be found explicitly
is very limited \cite{BLM}. We show that the algebraic properties of
the $ \mbox{$\hat{x}^{(n)}_{\theta}$}_{\cal H} $ 
differ significantly from those of the powers of
$\hat{x}_{\theta}$. In particular 
$\mbox{$\hat{x}^{(2)}_{\theta}$}_{\cal H} 
\neq (\mbox{$\hat{x}^{(1)}_{\theta}$}_{\cal H})^{2}$. 
This property will have immediate
consequences in the discussion of the uncertainty relation with
imperfect detectors. 

This paper has the following structure.
First, in Sec.~\ref{Sec:Zlambda}, we derive the POVM and the
generating operator 
for the operational observables. Their explicit form is found in the
limit of a classical local oscillator in Sec.~\ref{Sec:xthetaH}.
Given this result, we discuss the operational uncertainty relation
in Sec.~\ref{Sec:Uncertainty}. 
In Sec.~\ref{Sec:Random} we derive the family of operational homodyne
observables for the homodyne detector with a random phase between
the signal and the local oscillator fields, and relate them to the
intrinsic photon number operator. These calculations link the homodyne
noise with fluctuations of the photon statistics, and can be useful in
the time--resolved measurement of the properties of pulsed diode lasers. 
Finally, Sec.~\ref{Sec:TheEnd} summarizes the results.

\section{Generating operator for homodyne detection}
\label{Sec:Zlambda}
The family of the operational homodyne quantum observables
defined in Eq.~(\ref{operadef}) 
can be written conveniently with the help of the generating operator
\begin{equation}
\label{Eq:ZHlambdaDef}
\hat{Z}_{\cal H} (\lambda) = \int \text{d}a\,e^{i\lambda a}
\hat{\cal H}(a).
\end{equation}
Operational quantum observables are given by derivatives
of the generating operator at $\lambda = 0$:
\begin{equation}
\hat{x}^{(n)}_{\cal H} = \left. \frac{1}{i^n} 
\frac{{\text d}^n}{\mbox{\rm d}\lambda^n} 
\hat{Z}_{\cal H} (\lambda)
\right|_{\lambda=0}.
\end{equation}
This compact representation will noticeably 
simplify further calculations.
 
We will start the  calculations by finding
the generating operator for the homodyne detector. In a balanced setup,
the signal field 
described by an annihilation operator $\hat{a}$,
is superimposed on a local oscillator  $\hat{b}$ by means of a 50\%:50\%  
beamsplitter. The annihilation operators of the outgoing modes are 
given, up to the irrelevant phase factors, by the relation
\begin{equation}
\left( 
\begin{array}{c} \hat{c} \\ \hat{d} \end{array}
\right)
= 
\frac{1}{\sqrt{2}}
\left(
\begin{array}{rr} 1 & 1 \\ 1 & -1 \end{array}
\right)
\left(
\begin{array}{c} \hat{a} \\ \hat{b} \end{array}
\right).
\end{equation}
We will assume that the local oscillator is in a coherent state
 $|\beta\rangle_{LO}$. If another state of the local oscillator is
considered, our formulae can be generalized in a straightforward manner
by averaging the results over an appropriate Glauber's $P$-representation.

A quantity recorded in the experiment is the statistics
of the difference counts  between photodetectors facing the
modes $\hat{c}$ and $\hat{d}$. The difference of the counts $\Delta N$
corresponds to the classical variable, denoted before 
 as $a$, recorded in a homodyne detection experiment. 
The POVM $\hat{\cal H}(\Delta N)$ describing this
detection scheme can be easily derived.
It is clear that this POVM is an
operator acting in the Hilbert space of the signal mode.
Its explicit form can be found with the help of
standard theory of photodetection \cite{photodetection}:
\begin{eqnarray}
\hat{\cal H} (\Delta N) & = & \sum_{n_1 - n_2 = \Delta N}
\text{Tr}_{LO} \{ |\beta\rangle\langle\beta|_{LO}
\nonumber \\
& &
: e^{ -
\eta \hat{c}^\dagger \hat{c}} \frac{(\eta \hat{c}^\dagger \hat{c}
)^{n_1}}{n_1!} \, e^{ - \eta \hat{d}^\dagger \hat{d}} \frac{(\eta
\hat{d}^\dagger \hat{d})^{n_2}}{n_2!} : \}
\end{eqnarray}
where $\eta$ is the quantum efficiency, assumed to be identical for 
both the detectors. In this formula the  partial trace is over the local
oscillator mode and a marginal average with a fixed value 
of $\Delta N$ is performed. We will now convert this POVM into the 
generating operator according to Eq.~(\ref{Eq:ZHlambdaDef}).
As we discuss later, the POVM $\hat{\cal H}$ and consequently the 
generating operator $\hat{Z}_{\cal H}$ have their natural
parametrization, independent on the LO intensity. Before we find this
scaling, we will use $\xi$ instead of $\lambda$ as a parameter
of the generating operator:
\begin{eqnarray} 
\hat{Z}_{\cal H} (\xi) & = & \sum_{\Delta N = - \infty}^{\infty}
e^{i\xi \Delta N} \hat{\cal H} ( \Delta N)
\nonumber \\
& = & 
\text{Tr}_{LO} \{ |\beta\rangle\langle\beta|_{LO}
\nonumber \\
\label{ZHlambda1}
& &  : \exp [ \eta (e^{i\xi} - 1 )
\hat{c}^\dagger \hat{c} + \eta (e^{-i\xi} -1) \hat{d}^\dagger \hat{d}
] : \, \}.
\end{eqnarray}

Let us transform this expression to the form which does not
contain the normal ordering symbol. For this purpose we will use
the technique developed by Yuen and Shapiro \cite{YuenShapCQOIV} consisting
of extending the Hilbert space by two additional modes $\hat{c}_v$ and
$\hat{d}_v$ and constructing the fields annihilation operators
 \begin{equation}
\label{cvdv}
\hat{c}_{d} = \sqrt{\eta} \hat{c} + \sqrt{1-\eta} \hat{c}_v, \;\;\;\;\;
\hat{d}_{d} = \sqrt{\eta} \hat{d} + \sqrt{1-\eta} \hat{d}_v.
\end{equation}
The generating operator can be written in the extended four--mode space
using these operators as
\begin{eqnarray}
\label{ZHlambda2}
\hat{Z}_{\cal H} (\xi)
& = & \text{Tr}_{LO,v} \{ |\beta\rangle\langle\beta|_{LO} \otimes
|0 \rangle \langle 0 |_v
\nonumber \\
& & 
 : \exp[(e^{i\xi} - 1 )
\hat{c}_d^\dagger \hat{c}_d + (e^{-i\xi} -1) \hat{d}_d^\dagger \hat{d}_d
] : \, \},
\end{eqnarray}
where $\text{Tr}_v$ denotes the trace over both the vacuum modes 
$\hat{c}_v$ and $\hat{d}_v$. We can now apply the relation \cite{NormOrdExp} 
\begin{equation}
\label{normalny}
 : \exp[(e^{i\xi} -1) \hat{v}^\dagger
\hat{v}]: \, = \exp(i\xi  \hat{v}^{\dagger} \hat{v})
\end{equation}
valid for an arbitrary bosonic 
annihilation operator $\hat{v}$, which
finally gives:
\begin{eqnarray}
\label{ZHlambda3}
\lefteqn{\hat{Z}_{\cal H} (\xi)} & & \nonumber \\
& = & \text{Tr}_{LO,v} \{ |\beta\rangle\langle\beta|_{LO} \otimes
|0 \rangle \langle 0 |_v \,
\exp[i\xi (\hat{c}_d^\dagger \hat{c}_d - 
\hat{d}_d^\dagger \hat{d}_d)]
 \}.
\nonumber \\
& & 
\end{eqnarray}
This expression contains the most compact form of the homodyne POVM. The
exponent in Eq.~(\ref{ZHlambda2}) resembles the one from 
Eq.~(\ref{ZHlambda1}), 
with $\hat{c},\hat{d}$ replaced by $\hat{c}_d, \hat{d}_d$ and 
the detector efficiency equal to one. It is known \cite{YuenShapCQOIV},
that there is a physical picture behind this similarity.
An imperfect photodetector can be equivalently 
described by an ideal detector 
preceded by a beamsplitter with the power
transmissivity equal to the quantum 
efficiency of the real detector,
assuming that the vacuum state enters
through the unused port of the beamsplitter.
Mathematically, this construction corresponds to the so called Naimark
extension  of the POVM into a projective measure on a larger Hilbert
space \cite{BLM}. 

\section{Approximation of  a classical local oscillator}
\label{Sec:xthetaH}
When the local oscillator is in a strong coherent state, the
bosonic operators $\hat{b}, \hat{b}^\dagger$ can be replaced by
$c$-numbers $\beta, \beta^\ast$. However this approximation
violates the bosonic commutation relations for the pairs
$\hat{c}, \hat{c}^\dagger$ and $\hat{d}, \hat{d}^\dagger$, which
have been used implicitly several times in the manipulations 
involving $\hat{Z}_{\cal H} (\xi)$.
Therefore some care should be taken when considering the
classical limit of the local oscillator.

We will perform the approximation on the exponent of
Eq.~(\ref{ZHlambda3}). We will replace the quantum
average over the state $|\beta\rangle$ by 
inserting $\beta, \beta^\ast$ in place of
$\hat{b}, \hat{b}^\dagger$ and keep only the terms linear in
$\beta$. This gives
\begin{equation}
\hat{c}_d^\dagger c_d - \hat{d}_d^\dagger \hat{d}_d =
\sqrt{\eta} \beta^{\ast} \left( \sqrt{\eta} \hat{a} + \sqrt{1-\eta}
\frac{\hat{c}_v + \hat{d}_v}{\sqrt{2}}
\right) + \mbox{\rm h.c.}
\end{equation}
The operator in the brackets has a form analogous to Eq.~(\ref{cvdv})
with the combination $(\hat{c}_v + \hat{d}_v)/\sqrt{2}$ as a vacuum mode.
Consequently, the imperfectness of the photodetectors 
in the balanced homodyne detection can be
modelled by superposing the signal on a fictitious vacuum
mode before superposing it with the local oscillator
and attenuating the amplitude of the local oscillator field by
$\sqrt{\eta}$. This observation has been originally 
made by Leonhardt and Paul \cite{LeonPaulPRA93}, and is an example of
the Naimark extension involving a nonquantized local oscillator. 

Under the approximation of a classical local oscillator, it is now easy
to perform the trace over the vacuum modes with the 
help of the Baker--Campbell--Hausdorff formula. This yields
\begin{equation}
\label{ZHlambdaClassLO}
\hat{Z}_{\cal H}(\xi) = \exp[-\xi^2
\eta (1- \eta) |\beta|^2/2 ] \exp [ i \xi \eta ( \beta \hat{a}^\dagger +
\beta^\ast \hat{a}) ].
\end{equation}
The exponent $\exp [ - \xi^2 \eta ( 1- \eta) |\beta|^2 /2 ]$
introduces a specific ordering of the
creation and annihilation operators in the generating operator.
Therefore the detector efficiency $\eta$ can be related to the ordering
of the operational observables. For example, for $\eta=1/2$ we 
get $\hat{Z}_{\cal H}(\xi) = \exp (i\xi\beta^{\ast} \hat{a}/2)
\exp (i \xi  \beta \hat{a}^\dagger/2)$, i.e.\ the generating 
operator is ordered antinormally.

The expansion of the generating operator into a power series
of $i\xi$ gives
\begin{equation}
\label{ZExpansion}
\frac{1}{i^n} \left. \frac{\mbox{\rm d}^n \hat{Z}_{\cal H}}{{\mbox{\rm
d}} \xi^n} \right|_{\xi=0} = \left( \frac{1}{i}
\sqrt{\frac{\eta(1-\eta)}{2}} |\beta| \right)^n H_n \left( i
\sqrt{ \frac{\eta}{1-\eta}} \hat{x}_\theta \right),
\end{equation}
where $H_n$ denotes the $n$th Hermite polynomial and
$\hat{x}_\theta$ is the standard  quadrature operator
\begin{equation} 
\label{xtheta}
\hat{x}_\theta = \frac{e^{i\theta} \hat{a}^\dagger +
e^{- i \theta}  \hat{a}}{\sqrt{2}}
\end{equation}
expressed in terms of the creation and annihilation operators 
of the signal field and dependent on the local oscillator 
phase $\theta$ defined as $\beta = |\beta| e^{i\theta}$. 
In the terminology of the operational approach to quantum measurement,
 $\hat{x}_\theta$
is called an {\it intrinsic quantum observable}, since
it refers to internal properties of the system independent of
the measuring device \cite{EnglWodkPRA95}.

It will be convenient to change the parameter of
the generating operator in order to make the derivatives
(\ref{ZExpansion})  independent of the 
amplitude of the local oscillator.
A scaling factor which can be directly obtained from an 
experiment is the square root of
the intensity of the local oscillator field measured by the
photodetector. We will multiply it by $\sqrt{2}$
in order to get the intrinsic quadrature operator (\ref{xtheta})
in the limit $\eta \rightarrow 1$. Thus substituting 
$\xi=\lambda/\sqrt{2\eta|\beta|^2}$ yields the generating operator 
independent of the amplitude of the local oscillator:
\begin{equation}
\hat{Z}_{\cal H} (\lambda, \theta) = \exp [-\lambda^2(1-\eta)/4 + 
i\lambda\sqrt{\eta/2}(e^{i\theta}\hat{a}^\dagger 
+ e^{-i\theta}\hat{a})].
\end{equation}
The derivatives of $\hat{Z}_{\cal H}(\lambda,\theta)$ 
give  the final form of the family of the
operational  
observables ${\mbox{$\hat{x}$}_\theta^{(n)}}_{\cal H}$
for the homodyne detector:
\begin{equation}
{\mbox{$\hat{x}$}_\theta^{(n)}}_{\cal H} =
\left( \frac{\sqrt{1-\eta}}{2i} \right)^n H_n \left( i \sqrt{
\frac{\eta}{1-\eta} } \hat{x}_\theta \right).
\end{equation}

The algebraic properties of
the operational observables are quite complicated, since
${\mbox{$\hat{x}$}_\theta^{(n)}}_{\cal H}$
is not simply an $n$th power of
${\mbox{$\hat{x}$}_\theta^{(1)}}_{\cal H}$. 
Thus a single operator does not suffice to describe the 
homodyne detection with imperfect detectors. Complete 
characterization of the setup is provided by the whole family of 
operational observables. In fact the operators
${\mbox{$\hat{x}$}_\theta^{(n)}}_{\cal H}$ define an infinite algebra of 
operational homodyne observables for an arbitrary state of the signal mode.
As  mentioned above, for $\eta=50\% $, the general formula reduces
to antinormally ordered powers of the intrinsic quadrature operators:

\begin{equation}
{\mbox{$\hat{x}$}_\theta^{(n)}}_{\cal H}  
=  \frac{1}{2^{n/2}}
\vdots (\hat{x}_\theta)^{n}\vdots \ .
\end{equation}
This expression shows that the operational operators are  in some sense
equivalent to a prescription  of ordering of the intrinsic operators. This
prescription is dynamical in character, i.e., it depends on the efficiency
$\eta$ of the detectors used in the homodyne detection. In fact the homodyne
operational algebra is defined by a one-parameter family of dynamical
orderings defined by the generating operator derived in this section.

\section{Operational uncertainty relation}
\label{Sec:Uncertainty}
With  explicit forms of operational observables in hand,
we can now analyze their relation to the intrinsic quadrature
operator. For this purpose, 
let us look at the first lowest--order
operational quadrature observables:
\begin{eqnarray}
{\mbox{$\hat{x}$}_\theta^{(1)}}_{\cal H} & = & \eta^{1/2}
\hat{x}_\theta, \nonumber \\
{\mbox{$\hat{x}$}_\theta^{(2)}}_{\cal H} & = & \eta \left(
\hat{x}_\theta^2 + \frac{1 - \eta}{2\eta} \right), \nonumber \\
{\mbox{$\hat{x}$}_\theta^{(3)}}_{\cal H} & = & \eta^{3/2}
\left( \hat{x}_\theta^3 + \frac{3}{2} \frac{1-\eta}{\eta} 
\hat{x}_\theta \right).
\end{eqnarray}
The imperfectness of photodetectors influences the operational
observables in two ways. The first one is a trivial rescaling of the
observables by the powers of $\sqrt{\eta}$, the second way is a
contribution of the lower--order terms to the operational
counterparts
of $\hat{x}_\theta^n$.
In order to see its consequences
let us investigate the rescaled operational variance 
$\overline{(\Delta N)^{2}}-(\overline{\Delta N})^{2}$:
\begin{equation}
{\delta x_\theta^2}_{\cal H} = 
\frac{1}{2\eta|\beta|^2}
\left(\overline{(\Delta N)^{2}}-
(\overline{\Delta N})^{2}\right)
\end{equation}
From the definitions
of the operational operators it
is clear that this operational variance
involves ${\mbox{$\hat{x}$}_\theta^{(2)}}_{\cal H}$ and 
${\mbox{$\hat{x}$}_\theta^{(1)}}_{\cal H}$. The combination of these two
operators is in general different from the intrinsic variance. 
Because of this the operational dispersion of
$x_\theta$ is:
\begin{equation}
\label{deltaxthetaH}
{\delta x_\theta^2}_{\cal H} =
\langle {\mbox{$\hat{x}$}_\theta^{(2)}}_{\cal H} \rangle -
\langle {\mbox{$\hat{x}$}_\theta^{(1)}}_{\cal H} \rangle^2
= \eta \left( \Delta x _\theta^2 + \frac{1-\eta}{2\eta}
\right),
\end{equation}
where $\Delta x_\theta = \sqrt{\langle 
\hat{x}_\theta^2 \rangle - \langle
\hat{x}_\theta \rangle^2}$ is the intrinsic quantum dispersion of
the quadrature $x_\theta$. This intrinsic dispersion is
enhanced by a term coming from the imperfectness of the detectors.
Thus, the imperfectness of the photodetectors 
introduces an additional noise
to the measurement and deteriorates its resolution. 

Using the above
result we can derive the operational uncertainty relation for 
the quadratures related to the angles $\theta$ and $\theta'$ 
\begin{equation} 
\label{OperUncRel}
{\delta x_\theta}_{\cal H} {\delta x_{\theta'}}_{\cal H} \ge 
\eta \left( {\Delta x_\theta} {\Delta x_{\theta'}} + 
\frac{1-\eta}{2\eta} \right). 
\end{equation}
Again, an additional term is added to the intrinsic uncertainty 
product. This situation is similar to that in Ref.\ \cite{WodkPLA87} where
it was argued that taking into account the measuring device 
raises the minimum limit for the uncertainty product. However,
that discussion concerned a {\it simultaneous} measurement of
canonically conjugate variables, which is not the case in homodyne
detection. Using the intrinsic uncertainty relation 
$\Delta x_{\theta} \Delta x_{\theta'} \ge |\sin(\theta - \theta')|/2$
we get the result that the right hand side in the operational relation
(\ref{OperUncRel}) is not smaller than $(\eta|\sin(\theta-\theta')| 
+1-\eta)/2$.

One may wonder if the definition of squeezing is affected by the operational
operators. Let us consider the two quadratures ${\delta x_\theta}_{\cal H}$
and ${\delta x_{\theta+\frac{\pi}{2}}}_{\cal H}$. In this case the
operational uncertainty,
\begin{equation} 
{\delta x_\theta}_{\cal H} {\delta x_{\theta+\frac{\pi}{2}}}_{\cal H} \ge 
\frac{1}{2}, 
\end{equation}
is independent of $\eta$. However, it has to be kept in mind that only
a part of the operational dispersion comes from the field fluctuations.
The easiest way to discuss this is to 
rewrite Eq.~(\ref{deltaxthetaH}) to the form
\begin{equation}
{\delta x_\theta}_{\cal H} = \sqrt{\eta (\Delta x_\theta)^2
+ (1-\eta) \left( \frac{1}{\sqrt{2}} \right)^2 },
\end{equation}
which shows that the 
operational dispersion is a quadratic average of
the intrinsic field dispersion $\Delta x_\theta$
and the detector noise $1/\sqrt{2}$ that corresponds
to the vacuum fluctuation level. These contributions
enter with the weights $\eta$ and $1-\eta$,
respectively. Therefore if a squeezed quadrature
is measured with imperfect detectors, the
observed dispersion is larger than the intrinsic one.

\section{Homodyne detection with random phase}
\label{Sec:Random}
Homodyne detection is used primarily to detect phase--dependent
properties of light. However, it has been recently shown that even a
setup with a random phase between the 
signal and local oscillator fields can
be a useful tool in optical experiments \cite{MunrBoggPRA95}.
Although in this
case the phase sensitivity is lost, the homodyne detector can be
applied to measure phase--independent quantities and such a setup
presents some advantages over a single photodetector. First, the
information on the statistics of the  field is carried by the
photocurrent difference between the two rather intense fields. Within
existing detector 
technology, this quantity can be measured with a significantly better
efficiency than the weak field itself. Secondly, the spatio--temporal
mode that is actually measured by the homodyne detector 
is defined by the shape
of the local oscillator field. Consequently, application of 
the local oscillator in
the form of a short pulse allows the measurement to be performed with an
ultrafast sampling time. This technique
has been used in Ref.\ \cite{MunrBoggPRA95} to measure the time
resolved photon number statistics 
from a diode laser operating below threshold. The achieved
sampling time was significantly shorter
than those of previously used methods. 

The photon number distribution and other phase--independent quantities are
reconstructed from the average of the random phase homodyne statistics
calculated with the so--called {\it pattern functions}
\cite{LeonPaulPRA95}.  
For commonly used quantities, such as the diagonal elements of the density
matrix in the Fock basis, these pattern functions
take a quite complicated form. In this
section we will consider observables that are related to the
experimental data in the most direct way, the moments of the 
homodyne statistics with  randomized phase. 
We will derive the family of operational
observables and relate them to the powers of the photon number
operator $\hat{n} = \hat{a}^\dagger \hat{a}$. 

The generating operator for homodyne detection with random phase 
$\hat{Z}_{\cal R}$ ($\cal R$  stands for the random phase) is
obtained readily from $\hat{Z}_{\cal H}$ by averaging it over the
phase $\theta$. This gives
\begin{equation}
\label{Eq:ZRNormOrd}
\hat{Z}_{\cal R} (\lambda)  = 
\int_0^{2\pi} \frac{\text{d}\theta}{2\pi}
\hat{Z}_{\cal H} (\lambda, \theta) 
 =  e^{-\lambda^2/4} 
: J_0 \left(\lambda \sqrt{2\eta \hat{a}^\dagger \hat{a}} 
\right):  \ ,
\end{equation}
where $J_0$ is the Bessel function of the 0th order. 
With the help of the result derived in the Appendix, 
the normally ordered form of the Bessel function 
can be transformed into the following expression:
\begin{equation}
\hat{Z}_{\cal R} (\lambda)  =  e^{-\lambda^2/4} L_{\hat{n}}
(\eta\lambda^2/2), 
\end{equation}
where the index of the Laguerre polynomial is the the photon number
operator. The Laguerre polynomials with a operator valued index is
defined by the decomposition in the Fock basis. 

The family of operational observables is given by the derivatives
of the generating operator
\begin{equation}
\hat{x}^{(n)}_{\cal R} = \left. \frac{1}{i^n} 
\frac{{\text d}^n}{\mbox{\rm d}\lambda^n} 
\hat{Z}_{\cal R} (\lambda)
\right|_{\lambda=0}.
\end{equation}
Since the homodyne statistics averaged over the phase is even,
the odd derivatives disappear. A straightforward calculation yields
the operators for even $n=2m$: 
\begin{eqnarray}
\hat{x}_{\cal R}^{(2m)} & = & \frac{(2m-1)!!}{2^m} : L_m ( -2\eta
\hat{a}^{\dagger} \hat{a}): \nonumber \\
& = & \frac{(2m-1)!!}{2^m} \sum_{k=0}^m 
\left( \begin{array}{c} m \\ k \end{array} \right)
\frac{(2\eta)^k}{k!} 
\nonumber \\
& & \;\;\;\;\;\;\;\;\; \times
\hat{n} (\hat{n}-1)\ldots(\hat{n}-k+1).
\end{eqnarray}
This formula shows that 
$\hat{x}_{\cal R}^{(2m)}$
is a polynomial of $\hat{n}$
of the order of $m$. Therefore the first $m$ moments of the photon
number distribution can be computed from 
$\langle\hat{x}_{\cal R}^{(2)}\rangle,
\ldots,
\langle\hat{x}_{\cal R}^{(2m)}\rangle$. 
The two lowest--order observables
are given explicitly by 
\begin{eqnarray}
\hat{x}_{\cal R}^{(2)} & = & \eta \hat{n} + \frac{1}{2} 
\nonumber \\
\hat{x}_{\cal R}^{(4)} & = & \frac{3}{2} \left(
\eta^2 \hat{n}^2 + \eta ( 2- \eta) \hat{n} + \frac{1}{2} \right). 
\end{eqnarray}
It is seen that even in the case of ideal noise-free detectors 
$\hat{x}_{\cal R}^{(4)} \neq (\hat{x}_{\cal R}^{(2)})^2$ and the
family of the operational observables has nontrivial algebraic
properties. Inversion of the above equations yields:
\begin{eqnarray}
\hat{n} & = & \frac{1}{\eta} 
\left(\hat{x}_{\cal R}^{(2)} - \frac{1}{2} \right)
\nonumber \\
\hat{n}^2 & = & \frac{1}{\eta^2} 
\left( \frac{2}{3} \hat{x}_{\cal R}^{(4)} - (2-\eta)
\hat{x}_{\cal R}^{(2)} + \frac{1-\eta}{2} \right).
\end{eqnarray}
As an illustration, let us express the normalized photon number 
variance $Q = (\langle \hat{n}^2 \rangle 
- \langle \hat{n} \rangle^2 - \langle \hat{n} \rangle)/
\langle \hat{n} \rangle$ \cite{MandOL79} in terms of the 
expectation values of $\hat{x}_{\cal R}^{(2)}$ and
$\hat{x}_{\cal R}^{(4)}$. This variance is used to characterize the
sub-Poissonian statistics of light. After some simple
algebra we arrive at
\begin{equation}
Q = \frac{1}{\eta} 
\frac{\frac{2}{3}\langle\hat{x}_{\cal R}^{(4)}\rangle
- \langle\hat{x}_{\cal R}^{(2)}\rangle^2
-\langle\hat{x}_{\cal R}^{(2)}\rangle + \frac{1}{4}}%
{\langle\hat{x}_{\cal R}^{(2)}\rangle - \frac{1}{2}}.
\end{equation}
Thus, the variance $Q$ can be read out from the two lowest
moments of the homodyne statistics with the randomized phase. 
The photodetector efficiency $\eta$ enters into the above formula
only as an overall scaling factor. This result is analogous to that
obtained for the setup with a single imperfect detector, and is due to
the fact that $Q$ describes normally ordered field fluctuations. 

\section{Conclusions}
\label{Sec:TheEnd}
We have presented the operational description 
of the balanced homodyne detection
scheme with imperfect photodetectors. For homodyne detection it is
possible to derive exact expressions for the POVM and the corresponding
algebra of operational operators. The result of these calculations
shows  that a whole family of operational observables rather than a single
operator should to be used to discuss a realistic setup. This family
allows one to easily relate the experimentally observed 
fluctuations to the intrinsic properties of the system.

\section*{Acknowledgments}
The authors have benefited from 
discussions with P. L. Knight and G. Herling. 
This work was partially supported by
the US Air Force Phillips 
Laboratory Grant No. F29601-95-0209.
K.B. thanks the European Physical
Society for the EPS/SOROS
Mobility Grant.

\appendix

\section*{}

In this Appendix we present details of the transformation
of the generating operator $\hat{Z}_{\cal R} (\lambda)$
into the form that does not contain the normal ordering symbol. 
Let us start by rewriting Eq.~(\ref{normalny}) to the form
$\exp(\varepsilon \hat{a}^{\dagger} \hat{a})
=(1+\varepsilon)^{\hat{n}}$ and 
decomposing its right hand side of  in the Fock 
basis $\{ |n\rangle\}$:
\begin{eqnarray}
\label{Eq:Expansion}
:e^{\varepsilon \hat{a}^\dagger \hat{a}}: \, 
& = & 
\sum_{n=0}^{\infty} (1 + \varepsilon)^{n}
|n\rangle\langle n|
\nonumber \\
& = &
\sum_{n=0}^{\infty} \sum_{k=0}^{n} 
\left( \begin{array}{c} n \\ k \end{array} \right)
\varepsilon^{k}
| n \rangle \langle n |
\nonumber \\
& = &
\sum_{k=0}^{\infty} \varepsilon^{k} \sum_{n=k}^{\infty} 
\left( \begin{array}{c} n \\ k \end{array} \right)
| n \rangle \langle n |.
\end{eqnarray}
If we now assume the convention that $\left(
\begin{array}{c} n \\ k \end{array}
\right) = 0$ for $n<k$, the range of the second sum
can be extended to $k=0$ to $\infty$. It is then natural 
to denote it as a binomial coefficient of the operator 
$\hat{n}$. 
Comparing the
equal powers of $\varepsilon$ in Eq.~(\ref{Eq:Expansion}) 
yields a very compact representation of  the normally
ordered powers of $\hat{a}^{\dagger}\hat{a}$ 
in terms of $\hat{n}$:
\begin{equation}
(\hat{a}^\dagger)^k \hat{a}^k = k!
\left(
\begin{array}{c} \hat{n} \\ k \end{array}
\right).
\end{equation}
Expanding the normally ordered Bessel function in
Eq.~(\ref{Eq:ZRNormOrd}) and applying the above identity gives:
\begin{eqnarray}
:J_0 \left( \lambda \sqrt{2\eta\hat{a}^\dagger \hat{a}}
\right) : 
& = &
\sum_{k=0}^{\infty} \frac{(-\eta\lambda^2/2)^k}{(k!)^2}
(\hat{a}^\dagger)^k \hat{a}^k \nonumber \\
& = & \sum_{k=0}^{\infty} 
\left( \begin{array}{c} \hat{n} \\ k \end{array} \right)
\frac{(-\eta\lambda^2/2)^k}{k!} \nonumber \\
& = & L_{\hat{n}}(\eta\lambda^2/2),
\end{eqnarray}
where the Laguerre polynomial with 
the photon number operator index
\begin{equation}
L_{\hat{n}}(x) = \sum_{n=0}^{\infty} 
L_{n} (x) |n\rangle\langle n |
\end{equation}
is defined analogously to the binomial
coefficient via decomposition in the Fock basis.

\end{document}